\documentclass[twocolumn,floatfix,eqsecnum]{revtex4-2}
\usepackage{graphicx}
\usepackage{amsmath}
\usepackage{amssymb}
\usepackage{bm}
\usepackage{hyperref}

\usepackage{color}

\newif\ifmarked
\markedfalse
\definecolor{revblue}{rgb}{0.0,0.2,0.8}
\ifmarked
\newcommand{\rev}[1]{\textcolor{revblue}{#1}}
\newenvironment{revpart}{\color{revblue}}{}
\else
\newcommand{\rev}[1]{#1}
\newenvironment{revpart}{}{}
\fi

\begin{document}

\author{C. W. J. Beenakker}
\title{Phonon Scharnhorst effect: Acoustic analogue of the\\
vacuum-fluctuation-boosted speed of light}

\affiliation{Instituut-Lorentz, Universiteit Leiden, P.O. Box 9506, 2300 RA Leiden, The Netherlands}

\date{November 2026}

\begin{abstract}
The Scharnhorst effect in quantum electrodynamics predicts a superluminal velocity shift $\delta c>0$ at low photon frequencies due to vacuum fluctuations between two mirrors. Although theoretically robust (no causality violation), the effect is extraordinarily small ($\delta c/c\approx 10^{-32}$ for mirrors $1\,\mu{\rm m}$ apart) and outside current experimental reach. Here we propose a condensed matter analogue, replacing the photons in the electromagnetic vacuum by phonon excitations of a Bose-Einstein condensate. In an idealized geometry, the two-dimensional surface of a cylinder (circumference $L$) at zero temperature, the axial speed of sound is increased due to zero-point fluctuations of the condensate by $\delta c/c\simeq 5\times 10^{-3}\,\tilde{g}$ at $L\simeq 2\,\xi$, to first order in the dimensionless interaction constant $\tilde{g}\simeq 10^{-1}$ of the Bose gas of healing length $\xi\simeq 1\,\mu{\rm m}$. This is 28 orders of magnitude larger than the photonic effect \rev{at a comparable confinement length}, basically because phonons have a self-interaction while photons only interact via massive electrons. \rev{When $L$ drops below $1.19\,\xi$ the velocity shift changes sign, signaling the crossover to a one-dimensional Bose gas.} We discuss the prospects for observation of the acoustic Scharnhorst effect\rev{. The main obstacles are the nK temperature requirement and the need to separate the vacuum-fluctuation shift from a mean-field geometric correction.}
\end{abstract}
\maketitle

\section{Introduction}
Condensed-matter physics allows for the study of low-energy analogies of effects from high-energy physics. Examples include Klein tunneling of relativistic Dirac fermions \cite{Kle29}, observed as inter-band tunneling in graphene \cite{Kat06,Bee08,All11}, Schwinger particle-hole pair production \cite{Sch51}, also observed in graphene \cite{Sch23}, and the analogue of Hawking radiation \cite{Haw74} in Bose--Einstein condensates \cite{Ste16}. Such condensed-matter analogues provide both conceptual insight and experimental access to otherwise inaccessible high-energy processes. Here we contribute to this research programme by studying the acoustic analogue of an effect from quantum electrodynamics (QED): The superluminal velocity shift of light due to vacuum fluctuations in a confined geometry first predicted by Klaus Scharnhorst in 1990 \cite{Sch90,Bar90,Bar93,Lat95,Dit98,Lib01,Ish22}.

The Scharnhorst effect has been called ``one of the most elusive phenomena in modern physics'' \cite{Ani26}, both because it appears to be unmeasurably small (a relative velocity increase $\delta c/c\approx 10^{-32}$ in the vacuum between two mirrors $L=1\,\mu\text{m}$ apart \cite{note0}) and also because faster-than-$c$ propagation needs to be reconciled with causality (which it has been \cite{Mil90,Ben90,Lib02}). 

The effect is so small \cite{Sch90},
\begin{equation}
\delta c/c=\tfrac{11}{8100}\,\pi^2\alpha^2(\lambda_{\rm C}/L)^4,\label{QEDresult}
\end{equation}
because the photon-photon interaction in QED contributes to second order in the coupling constant $\alpha\approx \tfrac{1}{137}$ \cite{Greiner} and because it is mediated by massive electrons (Compton wave length $\lambda_{\rm C}=\hbar/m_{\rm e}c\approx 386\,{\rm fm}\ll L$).

Superfluid phonons offer an analogue in which both suppressions are mitigated. The vacuum-fluctuation shift of the sound velocity in a Bose-Einstein condensate appears already to first order in the coupling constant $\tilde{g}$ of the atomic contact interaction, and the interaction is mediated by the massless phonons themselves: The role of $\lambda_{\rm C}$ is taken over by the healing length $\xi\simeq 1\,\mu{\rm m}$. 

The large healing length also introduces a complication: Surfaces lead to a non-uniformity of order $\xi/L$ of the superfluid density that reduces the sound velocity, overwhelming the quantum enhancement. This, we believe, is why the analogue of the Scharnhorst effect was not noticed in existing studies of vacuum fluctuations in confined Bose-Einstein condensates \cite{Ant04,Har05,Ede06,Sch09,Bis16,Zin18,Ilg18,Son22,Tod24}.

Here we avoid this complication by considering a condensate on the surface of a cylinder, effectively introducing a periodic boundary condition that confines at uniform density. The speed of sound is probed along the cylinder axis, where the momentum is continuous and the long-wavelength limit is accessible. We find a zero-temperature shift of the sound velocity $\delta c/c\simeq 5\times 10^{-3}\,\tilde{g}$ at $L\simeq 2\,\xi$. We discuss the experimental requirements, which are demanding \rev{both} in temperature \rev{and in accuracy, because the vacuum-fluctuation shift needs to be separated from a larger mean-field background}.

\section{Two-dimensional Bose gas on a cylinder}
We consider a Bose gas confined to a two-dimensional (2D) strip of width $L$ in the $x$--$y$ plane, infinitely long in the $x$-direction, with periodic boundary conditions connecting $y=0$ to $y=L$. This corresponds to the surface of a cylinder of circumference $L$, if we neglect the curvature of the surface.

We seek the phonon velocity $c$ for propagation of a density wave (first sound) in the $x$-direction, at temperature $T=0$ and for wave lengths long compared to $L$. This is related to the compressibility by \cite{Pitaevskii}
\begin{equation}
mc^2=n\,d\mu/dn=n\,d^2 E/dn^2,\label{mc2relation}
\end{equation}
where $m$ is the atomic mass, $n$ the 2D density, $\mu$ the equilibrium chemical potential, and $E$ the ground state energy density of the condensate. \rev{The density derivatives are taken at fixed $L$ (or for the three-dimensional shell considered later: at fixed radius and fixed transverse confinement, see App.\ \ref{appA3}). Since the gas is infinitely long in the axial direction, canonical and grand-canonical descriptions give the same result.}

We start from the Gross--Pitaevskii theory \cite{Pitaevskii} of a weakly interacting, dilute Bose gas with contact interaction potential $g\sum_{i<j}\delta(\bm{r}_i-\bm{r}_j)$ of dimensionless strength $\tilde{g}=mg/\hbar^2\ll 1$. The characteristic length scale for order parameter inhomogeneities is the healing length $\xi=\hbar/\sqrt{2mgn}$, which for $\tilde{g}\ll 1$ is large compared to the inter-atomic separation. The bulk sound velocity is $c_0=\sqrt{gn/m}$, for $E=\tfrac{1}{2}gn^2\Rightarrow \mu=gn$.

The Bogoliubov approximation describes small-amplitude density and phase fluctuations of the condensate, via the pseudo-Hermitian eigenvalue problem
\begin{equation}
\begin{split}
&H_{\rm B}\begin{pmatrix}
u\\
v
\end{pmatrix}=\varepsilon\begin{pmatrix}
u\\
v
\end{pmatrix},\;\;H_{\rm B}=\sigma_z H_{\rm B}^\dagger\sigma_z,\\
&H_{\rm B}=\begin{pmatrix}
\hbar^2 |\bm{k}|^2/2m+gn&gn\\
-gn&-\hbar^2 |\bm{k}|^2/2m-gn
\end{pmatrix}.
\end{split}
\end{equation}

Because of translational invariance along $x$ the momentum component $k_x\equiv k$ is a good quantum number. The component $k_y$ is quantized by the periodic boundary conditions, $k_y=2\pi p/L\equiv q_p$, $p\in\mathbb{Z}$. The key feature of the periodic boundary condition, in contrast to hard-wall confinement, is that the unperturbed density $n$ remains uniform. The non-uniformities of order $\xi/L$ that hard walls would generate are absent by symmetry. The eigenvalues $\pm\varepsilon(k,q_p)$ of $H_{\rm B}$ are then given by
\begin{equation}
\varepsilon(k,q_p)=\hbar c_0 \sqrt{k^2+q_p^2}\sqrt{1+\tfrac{1}{2}\xi^2(k^2+q_p^2)}.\label{dispersion}
\end{equation}

\section{Regularization of the vacuum fluctuations}
At zero temperature each mode contributes $\tfrac{1}{2}\varepsilon(k,q_p)$ to the ground state energy via vacuum fluctuations. The sum $\sum_p\varepsilon(k,q_p)$ diverges; as usual in the Casimir effect \cite{Casimir}, the vacuum energy density is regularized by subtracting the bulk value, where the sum over $q_p$ is replaced by an integral over $q$. For a given longitudinal momentum $k$, the regularized contribution of vacuum fluctuations to the ground state energy density is
\begin{equation}
\delta E(k)=\frac{1}{L}\sum_{p=-\infty}^\infty \tfrac{1}{2}\varepsilon(k,2\pi p/L)-\int_{-\infty}^\infty \frac{dq}{2\pi}\,\tfrac{1}{2}\varepsilon(k,q).\label{deltaEk}
\end{equation}
The Abel-Plana formula \cite{note1} reduces this to
\begin{equation}
\delta E(k)=-\frac{1}{\pi}\int_0^\infty dt\,\frac{\operatorname{Im}\varepsilon(k,it)}{e^{tL}-1}.\label{AbelPlana}
\end{equation}
The imaginary momentum $q=it$ is approached from $\operatorname{Re}q>0$, where $\varepsilon(k,q)$ is analytic. The full vacuum energy density then follows upon integration over the longitudinal momentum, $\delta E=(2\pi)^{-1}\int_{-\infty}^\infty dk\,\delta E(k)$.

Since the imaginary part of $\varepsilon(k,it)$ is only nonzero for $k^2<t^2<k^2+2/\xi^2$, the integral expression for $\delta E$ has a 2D strip as integration domain,
\begin{align}
\delta E={}&-\frac{\hbar c_0}{\pi^2}\int_{0}^\infty dk\,\int_k^{\sqrt{k^2+2/\xi^2}}dt\,\frac{\sqrt{t^2-k^2}}{e^{tL}-1}\nonumber\\
&\times\sqrt{1+\tfrac{1}{2}\xi^2(k^2-t^2)}.
\end{align}
This can be rewritten as a function of the ratio $\ell=L/\xi$,
\begin{equation}
\begin{split}
&\delta E=-\frac{\hbar c_0}{L^3}F(\ell),\\
&F(\ell)=\int_{0}^\infty da\int_0^{\sqrt{2}}db\,\frac{\ell^{3} b^2 \sqrt{1-b^2 /2}}{\pi^2 \sqrt{a^2+b^2}(e^{\ell \sqrt{a^2+b^2}}-1)},
\end{split}\label{deltaEintegral}
\end{equation}
plotted in Fig.\ \ref{fig_plot}.

The small-$\ell$ and large-$\ell$ asymptotics serve as a check,
\begin{equation}
\delta E=-\frac{\hbar c_0}{\pi}\times
\begin{cases}
\frac{1}{3}(L\xi^2)^{-1}&\text{if}\;\;L\ll\xi,\\
\frac{1}{2}\zeta(3)L^{-3}&\text{if}\;\;L\gg\xi,
\end{cases}
\end{equation}
because these connect to known physics: The Casimir energy of a 2D massless scalar field with a compactified dimension for $L\gg\xi$ \cite{Casimir} and the Lieb-Liniger quantum correction \cite{Lie63} to the ground state energy of a 1D Bose gas for $L\ll\xi$.

\section{Vacuum-fluctuation modified sound velocity}
The correction $\delta c=c-c_0$ to the sound velocity from the vacuum fluctuation contribution $\delta E=E-E_0$ to the condensate energy follows the compressibility relation \eqref{mc2relation},
\begin{equation}
\frac{\delta c}{c_0} = \sqrt{1+\Delta}-1=\tfrac{1}{2}\Delta+{\cal O}(\Delta^2),\;\;
\Delta=\frac{n}{mc_0^2}\frac{d^2 \delta E}{dn^2}.
\end{equation}

\begin{figure}[tb]
\centerline{\includegraphics[width=0.9\linewidth]{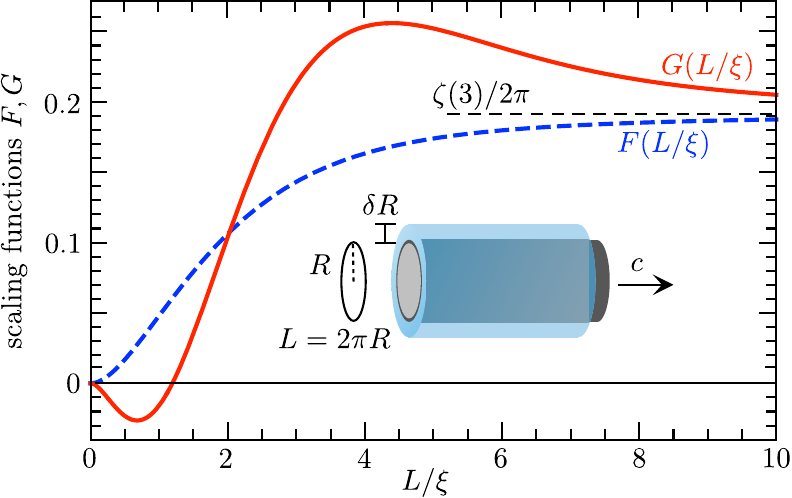}}
\caption{Scaling functions that determine the dependence on the ratio of cylinder perimeter $L$ and healing length $\xi$ of the vacuum-fluctuation contribution to the condensate energy [function $F$, Eq.\ \eqref{deltaEintegral}] and sound velocity [function $G$, Eq.\ \eqref{DeltaGeq}]. Both functions tend to $\zeta(3)/2\pi$ in the large-$L$ limit. The inset shows the cylinder geometry.
}
\label{fig_plot}
\end{figure}

In view of Eq.\ \eqref{deltaEintegral} the density dependence of $\delta E$ enters via $c_0=\sqrt{gn/m}$ and $\ell\equiv L/\xi = (L/\hbar)\sqrt{2mgn}$, resulting in
\begin{equation}
\Delta=\frac{\hbar c_0}{4gn^2 L^3}G(\ell),\;\;G(\ell)=F(\ell)-\ell F'(\ell)-\ell^2 F''(\ell).\label{DeltaGeq}
\end{equation}
The relative velocity shift thus becomes
\begin{equation}
\frac{\delta c}{c_0}=\frac{\sqrt{2}\tilde{g}\xi^3}{4L^3}G(L/\xi)+{\cal O}(\tilde{g}^2),\label{finalresult}
\end{equation}
with asymptotics
\begin{equation}
\frac{\delta c}{c_0}=\frac{\sqrt{2}\tilde{g}}{8\pi}\times\begin{cases}
\zeta(3)(\xi/L)^3&\text{if}\;\;L\gg\xi,\\
-2\xi/L&\text{if}\;\;L\ll\xi.
\end{cases}\label{asymptoticresult}
\end{equation}

Two features of the scaling function $G$ in Fig.\ \ref{fig_plot} are noteworthy. Firstly, the large-$L$ asymptotics \eqref{asymptoticresult} underestimates the shift at intermediate lengths: $G(L/\xi)$ is maximal at $L=4.39\,\xi$ where it exceeds the asymptotic value by $34\%$. The resulting velocity shift $\delta c/c_0\propto G(\ell)/\ell^3$ reaches a maximum $4.6\times 10^{-3}\,\tilde{g}$ at $L= 1.88\,\xi$. For $\tilde{g}\simeq 0.1$ this is a $5\times 10^{-4}$ effect, some 28 orders of magnitude above the QED Scharnhorst effect. Secondly, when $L$ drops below $1.19\,\xi$ the velocity correction changes sign. This is the crossover to the one-dimensional regime, where quantum fluctuations are known to \textit{reduce} the sound velocity \cite{Lie63}.

\begin{revpart}
The result \eqref{finalresult} is of first order in $\tilde{g}$. Corrections beyond the Bogoliubov approximation are of relative order $\tilde{g}$ (up to logarithmic factors), so at $\tilde{g}\simeq 0.1$ they modify the coefficients by an amount of order $10\%$, without changing the sign or the $L$-dependence of the effect. Phase fluctuations, which limit the validity of the Bogoliubov description in reduced dimensions, are governed by the Luttinger parameter $K=\pi\hbar nL/mc_0=\pi\ell/\sqrt{2}\tilde{g}$. At $\ell=1.88$ and $\tilde{g}=0.1$ one has $K=42$, deep in the weakly interacting quasi-condensate regime where Bogoliubov theory is accurate. In the one-dimensional limit $\ell\ll 1$ our result reproduces the weak-coupling expansion of the exact Lieb-Liniger solution \cite{Lie63}.
\end{revpart}

\section{3D effects}
The above calculation is for a strictly 2D Bose gas. The cylinder geometry is 3D, which introduces effects of curvature (radius $R=L/2\pi$) and finite transverse extension $\delta R$ of the condensate (see Fig.\ \ref{fig_plot}). This is a purely geometric effect if $\delta R$ is large compared to the scattering length $a_0$ of the Bose gas \rev{(see App.\ \ref{appA4} for corrections of higher order in $a_0/\delta R$)}. For harmonic confinement (frequency $\omega$), in the lowest mode ($\delta R=\sqrt{\hbar/2m\omega}$) we find that the curvature increases the coupling constant by a relative amount $\delta{g}/{g}=\tfrac{1}{2}(\delta R/R)^2$, see App.\ \ref{appA}.

The complication this presents is that the bare value $c_0$ now obtains an $R$-dependent correction $\delta c_0=\tfrac{1}{4}(\delta R/R)^2 c_0$, so it cannot be obtained from an independent measurement in a planar Bose gas. Moreover, $\delta c_0$ becomes weakly density dependent via a contribution of order $(\delta R/\xi)^2$, see App.\ \ref{appA}. \rev{Higher orders in $\delta R/R$ do not introduce a further density dependence: At fixed transverse confinement they are density independent and only add to the constant offset. The density dependence of $\delta c_0$ comes entirely from the interaction-induced deformation of the transverse profile, which is analytic in $n$ (App.\ \ref{appA3}). In addition, the effective 2D coupling constant has a weak logarithmic density dependence, which is characteristic of quasi-2D Bose gases (see App.\ \ref{appA4}).} To minimize the 3D effects we would therefore need to work in the regime $a_0\ll\delta R\ll R\lesssim\xi$.

\section{Finite temperature}

The $T=0$ velocity shift computed so far is entirely due to vacuum fluctuations in the condensate. When $T\neq 0$ thermal excitations enter, predominantly of the gapless $q=0$ mode. The leading order correction to the velocity shift is of order $(k_{\rm B}T L/\hbar c_0)^2$, with a coefficient that depends on whether the isothermal, adiabatic, or collisionless sound velocity is probed, see App.\ \ref{appB}. (All three velocities become identical at $T=0$.)

The characteristic temperature $T^\ast=\hbar c_0/k_{\rm B}L=\hbar^2/(\sqrt{2}\,mk_{\rm B}\xi L)$ favors light atoms such as $^7$Li. Still, $T^\ast$ is in the nK range for $\xi$ and $L$ of order $1\,\mu{\rm m}$, requiring extrapolation to $T=0$ rather than a direct measurement (although temperatures of a few nK have been reported \cite{Olf15}). Reducing $L$ is helpful because it increases $T^\ast$, but it also makes the curvature corrections more pronounced.

\begin{revpart}
To be specific, for $^7$Li at $\xi=1\,\mu{\rm m}$ and $L=1.88\,\xi$ (the maximum of the velocity shift) one has $T^\ast=26\,{\rm nK}$. The full finite-temperature expression of App.\ \ref{appB} shows that the isothermal velocity shift decreases by $10\%$ at $\tau\equiv T/T^\ast=0.077$, so at $T=2\,{\rm nK}$, and it vanishes at $\tau=0.29$. The thermal correction to the adiabatic velocity has the opposite sign and a five times larger coefficient [Eq.\ \eqref{Tcoefficients}], reaching $10\%$ at $\tau=0.033$, so at $T=0.9\,{\rm nK}$. The axial sound modes considered below have frequencies in the $10\,{\rm Hz}$ range; for $T\ll T^\ast$ the thermal excitations form a dilute gas of axial phonons in a quasi-one-dimensional channel, with strongly suppressed collision rates, so we expect the collisionless regime to apply. The coefficient of the thermal correction is not known in that regime, but its $\tau^2$ scaling is, and we expect its magnitude to be comparable to that of the hydrodynamic coefficients. We conclude that the zero-temperature result requires $T\lesssim 1$--$2\,{\rm nK}$, or else an extrapolation to $T=0$ using the known $T^2$ dependence.
\end{revpart}

\section{Prospects for an experimental realization}
For $L\simeq 2\xi$ and $\xi\simeq 1\,\mu{\rm m}$ one needs to trap a Bose gas in a cylindrical shell of sub-micron radius. \rev{No existing platform provides such a shell, but several elements are available.} Atoms trapped in the evanescent field of a tapered optical nanofiber \cite{Vet10} \rev{are held at a distance of about $200\,{\rm nm}$ from a fiber of sub-micron radius, uniformly over millimeters of fiber --- however, in a one-dimensional optical lattice rather than as a continuous, azimuthally uniform condensate.} Radiofrequency-dressed shell (``bubble'') traps produce closed curved condensates \cite{Car22,Ton23}\rev{, so far on millimeter scales}, and the digital-micromirror box potentials that generate uniform quasi-2D gases \cite{Gau13,Nav21} supply the flat axial potential that makes the axial phonon spectrum discrete. \rev{At the density $n=1/2\tilde{g}\xi^2=5\,\mu{\rm m}^{-2}$ (for $\tilde{g}=0.1$, $\xi=1\,\mu{\rm m}$) a shell of circumference $L=1.88\,\mu{\rm m}$ holds about $10$ atoms per $\mu{\rm m}$ of axial length, so some $10^3$ atoms in a $100\,\mu{\rm m}$ long tube.}

In an axial box of length $L_{\rm ax}\gg L$ the standing-wave phonons have a frequency ladder $\omega_j=\pi j c/L_{\rm ax}$, resolved with percent accuracy in uniform Bose gases \cite{Gar19,Vil18,Chr21}; the Scharnhorst shift appears as a common relative displacement $\delta\omega/\omega=\delta c/c_0$ of the whole ladder. \rev{The Bogoliubov dispersion \eqref{dispersion} adds a relative shift $\tfrac{1}{4}(q_j\xi)^2$, with $q_j=\pi j/L_{\rm ax}$. This term is independent of $L$ and can be separated by its $j^2$ dependence; for $L_{\rm ax}=200\,\mu{\rm m}$ it is $6\times 10^{-5}$ for the lowest mode.}

The $R$-dependent velocity shift has both a contribution $\delta c$ from the vacuum fluctuations and a geometric contribution $\delta c_0$ from the curvature and finite transverse extension of the cylindrical shell. \rev{For a harmonic transverse confinement of width $\delta R_0=85\,{\rm nm}$ (radial trap frequency $\omega/2\pi=100\,{\rm kHz}$ for $^7$Li) and a radius $R=L/2\pi=0.30\,\mu{\rm m}$, the geometric contribution $\delta c_0/c_0\approx\tfrac{1}{4}(\delta R_0/R)^2=2.0\times 10^{-2}$ is $43$ times larger than the vacuum-fluctuation shift $4.6\times 10^{-4}$ (at $\tilde{g}=0.1$). Reducing it below the signal would require $\delta R_0\lesssim 0.04\,R$, hence a radial trap frequency in the MHz range. The geometric effect therefore needs to be separated rather than suppressed.}

\rev{The two contributions can in principle be distinguished via their very different density dependence, provided the density scan includes both the sign change and the maximum of the vacuum-fluctuation shift. The geometric velocity shift from App.\ \ref{appA},
\begin{equation}
\delta c_0/c_0=\tfrac{1}{4}(\delta R\rev{_0}/R)^2-\rev{3\eta}(\delta R\rev{_0}/\xi)^2,\label{dc0result}
\end{equation}
with $\eta=\ln[4/(2+\sqrt{3})]=0.069$, is to leading order a constant plus a term linearly decreasing in $n$ (via $\xi\propto 1/\sqrt n$), to which the quasi-2D coupling adds a term $\propto(\tilde{g}/8\pi)\ln n$ (see App.\ \ref{appA4}). The vacuum fluctuations, in contrast, give a velocity shift $\propto G(\ell)/\ell^3$ with $\ell\propto\sqrt{n}$, which changes sign at $\ell=1.19$, is maximal at $\ell=1.88$, and decays as $n^{-3/2}$ for large $\ell$}.

\begin{revpart}
The azimuthal uniformity of the shell is a further requirement. The periodicity of the boundary condition is topological, but the trapping potential may vary around the cylinder, $V(y)=V_1\cos(2\pi y/L)$. To second order in $V_1$ this increases the mean-field compressibility velocity by $\delta c/c_0=\tfrac{1}{4}(V_1/\mu)^2x(1+x)^{-3}$, with $x=2\pi^2\xi^2/L^2$ and $\mu=gn$ \cite{note3}. At $L=1.88\,\xi$ this equals $4.9\times 10^{-3}(V_1/\mu)^2$, which remains below the signal for $V_1\lesssim 0.3\,\mu$ (corresponding to $10\,{\rm nK}$ for $^7$Li at $\xi=1\,\mu{\rm m}$). The modulation $V_1=mgR=2.5\,{\rm nK}$ from gravity, for a horizontal cylinder axis, contributes $2.5\times 10^{-5}$. Moreover, since $x\propto 1/n$, this contribution is to leading order density independent.
\end{revpart}

\section{Conclusion}
We have identified a condensed-matter analogue of the Scharnhorst effect: The speed of sound in a Bose--Einstein condensate on a cylinder is increased, relative to the bulk, by the depletion of the zero-point fluctuations of the phonon vacuum on the compact dimension. The correspondence between the photonic velocity shift \eqref{QEDresult} and its phonon counterpart \eqref{asymptoticresult} can be summarized by $\alpha^2(\lambda_C/L)^4\rightarrow\tilde{g}\,(\xi/L)^3$: The fine structure constant is replaced by the atomic interaction parameter, and the Compton wavelength by the healing length. \rev{At confinement lengths of order $1\,\mu{\rm m}$ t}his replacement increases the size of the effect by some 28 orders of magnitude\rev{. The main obstacles} to experimental observation in a Bose gas \rev{are} the nK temperature requirement --- needed to avoid thermal excitations from overwhelming the vacuum fluctuations \rev{--- and the need to separate the vacuum-fluctuation shift from a larger mean-field geometric background, by means of its characteristic density dependence.}

\begin{revpart}
In closing, we note that the analogy with quantum electrodynamics has motivated this study, but it has its limitations. In QED the speed of light is a universal limiting velocity, while in the Bose gas the speed of sound is a non-universal thermodynamic quantity. Issues of causality do not appear in the latter context. There is also a geometric difference: In QED the velocity of light propagating parallel to the mirrors, perpendicular to the confining direction, is unmodified, as a consequence of Lorentz invariance in the unconfined directions. That does not apply to the Bose gas, which allows for a velocity shift along the axis of the cylinder.
\end{revpart}

\acknowledgments
I am indebted to Dario Grasso for drawing my attention to the Scharnhorst effect. Anton Akhmerov suggested the cylinder geometry. Klaus Scharnhorst gave helpful comments on the manuscript. I used AI (Claude models Opus \& Fable) as an interactive tool to explore the topic. My research is supported by the Netherlands Organisation for Scientific Research (NWO/OCW), as part of Quantum Limits (grant ID 10.61686/EBSOG21615).

\appendix

\section{Curvature correction to the quasi-2D coupling}
\label{appA}

In our analysis we have reduced the 3D cylinder geometry, a shell of radius $R$ and thickness $\delta R$ (see Fig.\ \ref{fig_plot}), to an effective 2D geometry with periodic boundary conditions. Here we compute the curvature correction $\propto (\delta R/R)^2$ to the effective 2D coupling constant $g_{\rm{2D}}$ (denoted simply by $g$ in the main text). For a similar calculation on a sphere, see Ref.\ \onlinecite{Ton22}.

\subsection{Geometric renormalization}
\label{appA1}
Consider a 3D dilute gas of bosons of mass $m$ and $s$-wave scattering length $a_0$. The Gross-Pitaevskii contact interaction density $g_{\rm{3D}}|\Psi|^4$ has coupling constant \cite{Pitaevskii} $g_{\rm{3D}} = 4\pi \hbar^2 a_0/m$. 

When the gas is confined to the 2D surface of a cylinder of radius $R$, the effective 2D coupling constant $g_{\rm 2D}$ is found by integrating out the radial degree of freedom $\rho=R+\zeta$, with measure $(\rho/R)d\rho=(1+\zeta/R)d\zeta$, weighted by the radial profile $|\psi(\zeta)|^4$,
\begin{equation}
    g_{\rm{2D}} = g_{\rm{3D}} \int|\psi(\zeta)|^4 (1+\zeta/R)\,d\zeta.
\end{equation}
This simple geometric renormalization assumes that the confinement width $\delta R\gg a_0$, to ensure that the atomic scattering remains 3D locally (see App.\ \ref{appA4}).

The radial profile $\psi(\zeta)$, normalized to
\begin{equation}
\int |\psi(\zeta)|^2(1+\zeta/R)\,d\zeta=1,
\end{equation}
is determined by the confining potential $V(\zeta)$ via the radial wave equation
\begin{equation}
-(\hbar^2/2m)\bigl[\psi''+\psi'/(R+\zeta)\bigr]+V(\zeta)\psi=\mu_\perp \psi.
\end{equation}
(We write $\mu_\perp$ for the chemical potential because it also contains the confinement energy, which the $\mu$ from the main text does not.)

The substitution $\psi=(1+\zeta/R)^{-1/2}\chi$, with normalization $\int|\chi|^2\,d\zeta=1$, removes the first-order-derivative term, producing a 1D Schr\"odinger equation with an additional curvature-induced potential \cite{Cos81},
\begin{equation}
\begin{split}
&-(\hbar^2/2m)\chi''+[\delta V(\zeta)+V(\zeta)]\chi=\mu_\perp\chi,\\
&\delta V(\zeta)=-\frac{\hbar^2}{8m(R+\zeta)^2}=-\frac{\hbar^2}{8mR^2}+{\cal O}(R^{-3}).
\end{split}
\label{1DSchrodinger}
\end{equation}
To order $1/R^2$ this potential only shifts the chemical potential $\mu_\perp$ by a density independent amount, so it can be ignored. Moreover, for a symmetric confinement, $V(-\zeta)=V(\zeta)$, the curvature does not break the symmetry $|\chi(-\zeta)|^2=|\chi(\zeta)|^2$ of the radial profile to order $1/R^2$.

The 2D coupling constant becomes,
\begin{align}
g_{\rm{2D}}={}&g_{\rm{3D}}\int \frac{|\chi(\zeta)|^4}{1+\zeta/R}\,d\zeta\nonumber\\
={}&g_{\rm{3D}}\int |\chi(\zeta)|^4[1+(\zeta/R)^2+{\cal O}(R^{-3})]\,d\zeta,
\label{g2chi}
\end{align}
assuming symmetric confinement (so the order $1/R$ contribution vanishes). In terms of the effective width,
\begin{equation}
\delta R^2=\int |\chi(\zeta)|^2\zeta^2\,d\zeta,
\end{equation}
we obtain
\begin{equation}
\begin{split}
&g_{\rm{2D}}=\frac{g_{\rm{3D}}}{\delta R}[C_0+C_2 (\delta R/R)^{2}+{\cal O}(\delta R/R)^{3}],\\
&C_n=\delta R^{1-n}\int |\chi(\zeta)|^4\zeta^n\,d\zeta.
\end{split}
\end{equation}
The numerical coefficients $C_n$ are $R$-independent, so independent of curvature. For harmonic ground-state confinement [frequency $\omega$, wave function $\chi_0\propto \exp(-\tfrac{1}{2}m\omega\zeta^2/\hbar)$], one has
\begin{equation}
 \delta R=\sqrt{\frac{\hbar}{2m\omega}},\;\;C_0=\frac{1}{2\sqrt{\pi}},\;\; C_2=\frac{1}{4\sqrt{\pi}},
\end{equation}
hence $\delta g/g=(C_2/C_0)(\delta R/R)^2=\tfrac{1}{2}(\delta R/R)^2$.

\subsection{Density dependence}
\label{appA2}
The quasi-2D description of the 3D Bose gas assumes that it occupies the lowest mode of the confining potential. The excess chemical potential $\delta\mu=\mu_\perp-\tfrac{1}{2}\hbar\omega$ should be small compared to the level spacing $\hbar\omega$. The interactions then act as a perturbation of the transverse profile, introducing a weak density dependence of $\delta R$ relative to the bare value $\delta R_0$, as we now calculate.

We include the contact interaction in the 1D Schr\"{o}dinger equation \eqref{1DSchrodinger}. To first order in the coupling constant we have, for harmonic confinement,
\begin{equation}
\begin{split}
&(H_0-\tfrac{1}{2}\hbar\omega)\delta\chi=(\delta\mu-U)\chi_0,\;\;U(\zeta)=g_{\rm{3D}}n|\chi_0(\zeta)|^2,\\
&H_0=-\frac{\hbar^2}{2m}\frac{d^2}{d\zeta^2}+\tfrac{1}{2}m\omega^2\zeta^2.
\end{split}
\end{equation}
Projection onto $\chi_0$ gives the expected first order correction to the chemical potential due to the contact interaction,
\begin{equation}
\delta\mu=g_{\rm{3D}}n\int |\chi_0|^4\,d\zeta=g_{\rm{2D}}n.
\end{equation}
The first order correction to the wave function is
\begin{equation}
\delta\chi=-\sum_{n\neq 0}\frac{|n\rangle\langle n|U|0\rangle}{n\hbar\omega},
\end{equation}
with $|n\rangle$ the $n$-th harmonic oscillator eigenstate.

The matrix element $\langle 0|\zeta^2|n\rangle$ is only nonzero for $n=0$ or $n=2$, hence
\begin{align}
\delta R^2={}&\langle\chi|\zeta^2|\chi\rangle=\langle\chi_0|\zeta^2|\chi_0\rangle+2\operatorname{Re}\langle\chi_0|\zeta^2|\delta\chi\rangle\nonumber\\
={}&\langle 0|\zeta^2|0\rangle-2\operatorname{Re}\frac{\langle 0|\zeta^2|2\rangle\langle 2|U|0\rangle}{2\hbar\omega}\nonumber\\
={}&\delta R_0^2 +\frac{g_{\rm{3D}}n}{4\sqrt{\pi}}\frac{\delta R_0}{\hbar\omega}.
\end{align}
This gives, with $\delta\mu=g_{\rm{2D}}n = g_{\rm{3D}}n(2\sqrt{\pi}\delta R_0)^{-1}$,
\begin{equation}
\delta R=\delta R_0\bigl[1+\tfrac{1}{4}\delta\mu/\hbar\omega\bigr]=\delta R_0\bigl[1+\tfrac{1}{4}(\delta R_0/\xi)^2].
\end{equation}

\subsection{\rev{Mean-field background of} the sound velocity}
\label{appA3}
\begin{revpart}
The density dependence of the transverse profile, found in App.\ \ref{appA2}, has to be included in the mean-field energy density $E_{\rm mf}(n)$, from which the sound velocity follows via the compressibility relation \eqref{mc2relation}. (Inserting a density-dependent coupling constant into $c_0^2=gn/m$ would not be correct.) 

Minimization of the Gross-Pitaevskii energy functional with respect to the transverse profile gives, to second order in the interaction,
\begin{equation}
\begin{split}
&E_{\rm mf}=n\varepsilon_\perp+\tfrac{1}{2}g_{\rm 2D}n^2-\eta\,g_{\rm 2D}^2n^3/\hbar\omega+{\cal O}(n^4),\\
&\eta=\hbar\omega\,\frac{\langle\chi_0^3|Q(H_0-\tfrac{1}{2}\hbar\omega)^{-1}Q|\chi_0^3\rangle}{\bigl(\int\chi_0^4\,d\zeta\bigr)^2},
\end{split}
\label{Emf}
\end{equation}
with $Q=1-|\chi_0\rangle\langle\chi_0|$ and $\varepsilon_\perp$ the density-independent transverse energy per particle (including the curvature-induced shift $-\hbar^2/8mR^2$). 

The cubic term, the energy gained by the relaxation of the transverse profile, is an effective three-body interaction \cite{Mer13}. It describes the same physics as the profile deformation of App.\ \ref{appA2}: By the Hellmann-Feynman theorem, $\partial E_{\rm mf}/\partial g_{\rm 3D}=\tfrac{1}{2}n^2\int\chi^4\,d\zeta$, so the cubic term corresponds to a relative change $-4\eta(\delta R_0/\xi)^2$ of $\int\chi^4\,d\zeta$. For harmonic confinement the sum over intermediate states can be carried out in closed form, with the result
\begin{equation}
\eta=\ln[4/(2+\sqrt{3})]=0.0693.
\end{equation}
(A Gaussian variational profile gives the approximation $\eta=1/16$.)

Substitution into Eq.\ \eqref{mc2relation} gives $mc_0^2=g_{\rm 2D}n(1-6\eta\,g_{\rm 2D}n/\hbar\omega)$, where $g_{\rm 2D}n/\hbar\omega=(\delta R_0/\xi)^2$. App.\ \ref{appA1} gives the curvature renormalization of $g_{\rm 2D}$, in dimensionless form
\begin{equation}
\tilde{g}=(2\sqrt{\pi}a_0/\delta R_0)[1+\tfrac{1}{2}(\delta R_0/R)^{2}].
\label{gtilde}
\end{equation}
Combining the two, we obtain the mean-field velocity shift relative to a flat layer,
\begin{equation}
\frac{\delta c_0}{c_0}=\frac{1}{4}\frac{\delta R_0^2}{R^2}-3\eta\frac{\delta R_0^2}{\xi^2}+{\cal O}\Bigl(\frac{\delta R_0^4}{R^4},\frac{\delta R_0^4}{R^2\xi^2},\frac{\delta R_0^4}{\xi^4}\Bigr).
\end{equation}
At fixed transverse confinement the first term, as well as its higher-order corrections in $\delta R_0/R$, is density independent, since it follows from the non-interacting transverse problem. The second term is linear in density, because $(\delta R_0/\xi)^2=2\tilde{g}n\,\delta R_0^2$. The neglected cross term is again linear in density, while the term of order $(\delta R_0/\xi)^4$ is quadratic in density but smaller by a factor $(\delta R_0/\xi)^2\approx 10^{-2}$ for the parameters of the main text.
\end{revpart}

\begin{revpart}
\subsection{Beyond the geometric coupling constant}
\label{appA4}

The geometric coupling constant $\tilde{g}=2\sqrt{\pi}\,a_0/\delta R$ of App.\ \ref{appA1} is the leading term in an expansion in $a_0/\delta R$. Corrections of next order, of relative order $\tilde{g}$, are of two types. Firstly, virtual transverse excitations produce an effective three-body interaction \cite{Mer13}, with a relative contribution of order $\tilde{g}n\,\delta R^2\simeq(\delta R/\xi)^2$. This is the interaction-induced deformation of the transverse profile, included in App.\ \ref{appA2} and \ref{appA3}. Secondly, the 2D scattering amplitude has a logarithmic energy dependence, which makes $\tilde{g}$ weakly density dependent, $d\ln\tilde{g}/d\ln n=\tilde{g}/4\pi$ \cite{Pet00}, hence $d\ln (c_0/\sqrt{n})/d\ln n=\tilde{g}/8\pi$.

Which of the two density dependencies dominates depends on the value of $n\delta R^2$. For the parameters of the experimental proposal ($n\delta R^2\approx 0.04$) their variations over the density range considered are comparable, of order $10^{-3}$--$10^{-2}$. Both enter in the mean-field background of the sound velocity, as smooth (linear and logarithmic) functions of density. Their effect on the vacuum-fluctuation shift \eqref{finalresult}, which is already of first order in $\tilde{g}$, is of higher order.
\end{revpart}

\section{Finite temperature}
\label{appB}

\subsection{Free energy} 
\label{appB1}
At non-zero temperature $T$ the zero-point vacuum energy $\tfrac{1}{2}\varepsilon$ of each mode is replaced by the free energy
\begin{equation}
f(\varepsilon,T) =\tfrac{1}{2}\varepsilon + T\ln(1-e^{-\varepsilon/T})\label{fdef}
\end{equation}
(setting Boltzmann's constant $k_{\rm B}$ to unity). The regularized sum over modes, for a given longitudinal momentum $k$, is then given by
\begin{align}
\delta E(k,T)={}&\left(\frac{1}{L}\sum_{p=-\infty}^\infty -\int_{-\infty}^\infty \frac{dq_p}{2\pi}\right)\,f[\varepsilon(k,q_p),T],
\end{align}
with $q_p=2\pi p/L$. \rev{(At nonzero temperature $\delta E$ thus denotes the regularized free energy density, rather than the ground state energy density; we retain the notation for continuity with the main text.)} This is analogous to the $T=0$ formula \eqref{deltaEk}, but the logarithm in $f$ changes the analytic structure, so the Abel-Plana formula \eqref{AbelPlana} needs to be reconsidered.

For that purpose we first rewrite Eq.\ \eqref{fdef} identically as $f=T\ln[2\sinh(\varepsilon/2T)]$. The infinite product formula $\sinh z=z\prod_{n=1}^\infty [1+(z/n\pi)^2]$ then gives $f$  as a sum over Matsubara frequencies $\omega_{n}=2\pi n T/\hbar$,
\begin{equation}
f(k,q,T)=\tfrac{1}{2}T\sum_{n=-\infty}^{\infty}
\ln\bigl[\varepsilon^{2}(k,q)+\hbar^{2}\omega_{n}^{2}\bigr]+\text{const}.
\end{equation}
The constant is a density-independent offset of the regularized free energy and can be ignored. 

If we substitute the dispersion relation \eqref{dispersion} for $\varepsilon(k,q)$ we may factorize,
\begin{equation}
\begin{split}
&\varepsilon^{2}(k,q)+\hbar^{2}\omega^{2}
=\tfrac{1}{2}\hbar^{2}c_{0}^{2}\xi^{2}
\bigl(q^{2}+\kappa_{+}^{2}\bigr)\bigl(q^{2}+\kappa_{-}^{2}\bigr),\\
&\xi^2\kappa_{\pm}^{2}(k,\omega)=\xi^2 k^{2}
+1\pm\sqrt{1-2\xi^{2}\omega^{2}/c_{0}^{2}}.
\end{split}
\end{equation}
\rev{For $2\xi^2\omega^2<c_0^2$ both $\kappa_\pm^2$ are real and non-negative; for $2\xi^2\omega^2>c_0^2$ they are complex conjugates with positive real part $k^2+\xi^{-2}$. In both cases we take the principal branch of the square root, so that $\operatorname{Re}\kappa_\pm\geq 0$ and the sum over $s=\pm$ in Eq.\ \eqref{Lifshitz} below is real.}
We now apply the Abel-Plana formula \cite{note1} to
\begin{align}
&\left(\frac{1}{L}\sum_{p=-\infty}^{\infty}-\int_{-\infty}^{\infty}\frac{dq_p}{2\pi}\right)
\ln\bigl(q_p^{2}+\kappa^{2}\bigr)=\\
&\quad=-\frac{2}{\pi}\int_0^\infty dt\,\frac{\operatorname{Im}\ln\bigl(\kappa^{2}-t^2)}{e^{tL}-1}=\frac{2}{L}\ln\bigl(1-e^{-\kappa L}\bigr), \nonumber
\end{align}
for $\operatorname{Re}\kappa>0$. This gives for the regularized free energy, omitting constants,
\begin{equation}
\delta E(k,T)=\frac{T}{L}\sum_{n=-\infty}^{\infty}\sum_{s=\pm}
\ln\bigl[1-e^{-\kappa_{s}(k,\omega_{n})L}\bigr].
\label{Lifshitz}
\end{equation}

Finally, we integrate over the longitudinal momentum to obtain the free-energy  scaling function $F$,
\begin{subequations}
\label{FlT}
\begin{align}
&\delta E=-(\hbar c_0/L^3)F(\ell,\tau),\;\;\ell=L/\xi,\;\; \tau=T L/\hbar c_0,\\
&F(\ell,\tau)=-\frac{\tau\ell}{\pi}\sum_{n=-\infty}^{\infty}\sum_{s=\pm}
\int_{0}^{\infty}da\nonumber\\
&\qquad\times\ln\Bigl[1-\exp\Bigl(-\ell\sqrt{a^{2}+1+s\sqrt{1-2\tilde{\nu}_{n}^{2}}}\Bigr)\Bigr],\\
&F(\ell,\tau)=F(\ell,0)+\frac{\pi}{6}\tau^2-\frac{\zeta(3)}{2\pi}\tau^3+{\cal O}(\tau^4),
\end{align}
\end{subequations}
with $\tilde{\nu}_{n}=2\pi n\tau/\ell$.

\subsection{Sound velocity}
\label{appB2}

\begin{figure}[tb]
\centering
\includegraphics[width=0.8\linewidth]{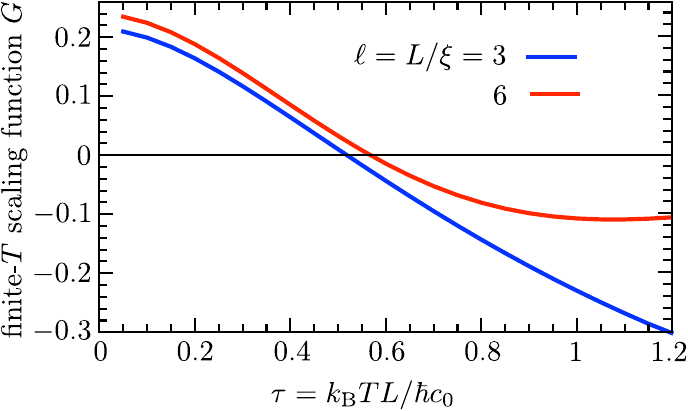}
\caption{Temperature dependence of the scaling function \eqref{GtauDef} that governs the $L$-dependent shift of the isothermal compressibility velocity $\delta c_T=\tfrac{1}{4}\sqrt{2}c_0\tilde{g}\ell^{-3}G(\ell,\tau)$.}
\label{fig:GT}
\end{figure}

The finite-temperature sound-velocity scaling function $G$ is given by
\begin{subequations}
\label{GtauDef}
\begin{align}
&\Delta=\frac{n}{mc_0^2}\frac{d^2 \delta E}{dn^2}=\frac{\hbar c_0}{4gn^2L^3}\,G(\ell,\tau),\\
&G=\bigl(1-\hat{D}^2\bigr)F,\;\;\hat{D}=\ell\,\partial_\ell-\tau\,\partial_\tau ,\\
&G(\ell,\tau)=G(\ell,0)-\frac{\pi}{2}\tau^2+\frac{4\zeta(3)}{\pi}\tau^3+{\cal O}(\tau^4),
\end{align}
\end{subequations}
Fig.\ \ref{fig:GT} shows $G(\ell,\tau)$ for two values of $L/\xi$.

Since the density derivatives in Eq.\ \eqref{GtauDef} are taken at fixed $T$, the scaling function $G(\ell,\tau)$ governs the shift $\delta c_T$ of the \textit{isothermal} compressibility velocity, 
\begin{equation}
\delta c_T/c_0=\tfrac{1}{2}\Delta=\frac{\sqrt{2}\tilde{g}\xi^3}{4L^3}G(\ell,\tau).
\end{equation}
A hydrodynamic density wave propagates at the \textit{adiabatic} velocity $c_{S}$, at fixed entropy per particle. Let us determine the corresponding shift $\delta c_S$.

Isothermal and adiabatic velocities are related by the thermodynamic identity \cite{Hei06,Fri18}
\begin{equation}
c_S^2-c_T^2=\frac{X^2}{m n\,Y},
\;\;
X=S-n\left(\frac{\partial S}{\partial n}\right)_{\!T},
\;\;
Y=\left(\frac{\partial S}{\partial T}\right)_{\!n},
\label{eq:cScT}
\end{equation}	
with $S$ the entropy per unit area. If $S\propto T^{a}n^{-b}$ then
\begin{equation}
c_S^2-c_T^2=a^{-1}(1+b)^2 ST/mn.
\label{eq:exponent}
\end{equation}
We now use that for temperatures $T\ll\hbar c_0/L$ the thermally populated excitations are predominantly those of the gapless $q=0$ mode, whose entropy per unit area is
\begin{equation}
S=\tfrac{1}{3}\pi T/\hbar c_0L,
\label{eq:sax}
\end{equation}
for a linear dispersion. [The nonlinearity introduces corrections of order $(\tau/\ell)^2$.] Since $c_0\propto n^{1/2}$ we have scaling exponents $a=1$, $b=1/2$,
\begin{equation}
c_S^2-c_T^2=\frac{9}{4}\,\frac{ST}{mn}=\frac{3\pi T^{2}}{4mn\hbar c_0L}.
\end{equation}

For the $L$-dependent shift we subtract the difference $[c_S^2-c_T^2]_\infty$ of a planar 2D Bose gas \cite{Pitaevskii}, when $a=2$, $b=1$,
\begin{equation}
[c_S^2-c_T^2]_\infty=\frac{2S_{\infty}T}{mn},\;\;S_\infty=\frac{3\zeta(3)}{2\pi}\frac{T^2}{(\hbar c_0)^2}.
\end{equation}
We thus arrive at
\begin{align}
&\delta (c_S^2-c_T^2)=(T/mn)(\tfrac{9}{4}S-2S_\infty)\nonumber\\
&\Rightarrow \delta c_S-\delta c_T=\frac{3\pi T^{2}}{8\,mn\,\hbar c_0^{2}L}
\left(1-\frac{4\zeta(3)}{\pi^{2}}\frac{TL}{\hbar c_0}\right).
\end{align}
At $L\gg\xi$ the leading order finite-temperature corrections are
\begin{equation}
\frac{\delta c}{c_0}=\frac{\sqrt{2}\tilde{g}\xi^3}{8\pi L^3}\times
\begin{cases}
\zeta(3)-\pi^2(TL/\hbar c_0)^2&\text{isothermal},\\
\zeta(3)+5\pi^2(TL/\hbar c_0)^2&\text{adiabatic}.
\end{cases}
\label{Tcoefficients}
\end{equation}

Both these velocities assume that the distribution of thermally excited quasiparticles re-equilibrates within a sound period. At low but nonzero temperatures this assumption breaks down and a collisionless quasiparticle velocity becomes the relevant quantity. The parametric dependence $\propto (TL/\hbar c_0)^2$ of the thermal correction to the velocity shift is fixed by the phase space of the gapless mode, but the coefficient will be different from the values in Eq.\ \eqref{Tcoefficients}.


\begin{thebibliography}{99}
\bibitem{note0} Many sources cite $10^{-36}$ rather than $10^{-32}$ for the size of the Scharnhorst effect at $L=1\,\mu{\rm m}$. This is based on a numerical error, see the Erratum of Ref.\ \onlinecite{Sch90}.
\bibitem{note1} The Abel-Plana formula $\bigl[\sum_{n=0}^\infty-\int_0^\infty dn\bigr]f(n)=\tfrac{1}{2}f(0)+i\int_0^\infty dt\,[f(it)-f(-it)](e^{2\pi t}-1)^{-1}$ refers to a one-sided sum, rather than to the two-sided sum we have here. To apply it, use that $\varepsilon(k,q)=\varepsilon(k,-q)\in\mathbb{R}$ for $q\in\mathbb{R}$, hence $L\delta E(k)=\bigl[\sum_{n=0}^\infty-\int_{0}^\infty dn\bigr]\varepsilon(k,2\pi n/L)-\tfrac{1}{2}\varepsilon(k,0)=i\int_0^\infty dt\,(e^{2\pi t}-1)^{-1}[\varepsilon(k,2\pi it/L)-\varepsilon(k,-2\pi it/L)]=-(L/\pi)\int_0^\infty dt\,(e^{tL}-1)^{-1}\operatorname{Im}\varepsilon(k,it)$.
\end{thebibliography}
\end{document}